\documentclass[twocolumn,aps,prb,showpacs,preprintnumbers,amsmath,aimssymb]{revtex4-1}
\usepackage{graphicx}
\usepackage{bm}
\usepackage{gensymb}
\usepackage{color}
\usepackage{url}

\begin{document}
\title{Effective magnetic correlations in hole-doped graphene nanoflakes}

\author{A.~Valli$^{1}$}
\author{A.~Amaricci$^{1}$}
\author{A.~Toschi$^{2}$}
\author{T.~Saha-Dasgupta$^{3}$}
\author{K.~Held$^{2}$}
\author{M.~Capone$^{1}$}

\affiliation{$^{1}$Democritos National Simulation Center, Consiglio Nazionale delle Ricerche,
Istituto Officina dei Materiali (IOM) and Scuola Internazionale Superiore di Studi Avanzati (SISSA), Via Bonomea 265, 34136 Trieste, Italy}
\affiliation{$^{2}$Institute for Solid State Physics, Vienna University of Technology, 1040 Wien, Austria}
\affiliation{$^3$S.~N.~Bose National Centre for Basic Sciences, 70098 Kolkata, India}




\date{\today}
\begin{abstract}
The magnetic properties of zig-zag graphene nanoflakes (ZGNF)
are investigated within the framework of the dynamical mean-field theory. 
At half-filling and for realistic values of the local interaction, 
the ZGNF is in a fully compensated antiferromagnetic (AF) state, 
which is found to be robust against temperature fluctuations. 
Introducing charge carriers in the AF background 
drives the ZGNF metallic and stabilizes a magnetic state 
with a net uncompensated moment at low temperature. 
The change in magnetism is ascribed to the delocalization 
of the doped holes in the proximity of the edges, 
which mediate ferromagnetic correlations between the localized magnetic moments. 
Depending on the hole concentration, the magnetic transition 
may display a pronounced hysteresis over a wide range of temperature, 
indicating the coexistence of magnetic states with different symmetry. 
This suggests the possibility of achieving the electrostatic control 
of the magnetic state of ZGNFs to realize a switchable spintronic device.  
\end{abstract}
\maketitle

\section{Introduction} 
Graphene is widely regarded as a promising material for nanoelectronics.\cite{snookGNF}  
The high electron mobility of the delocalized $\pi$-electrons 
in graphene results in excellent electric and thermal transport properties, 
leading the belief that graphene holds the potential to outperform Si 
for the realization of high-speed and high-frequency response transistors 
and large-scale integrated circuits with a low environmental impact. 
In this respect, the semi-metallic nature of graphene 
is not ideal for electronic applications, and represents 
the main limitation to the realization of a graphene transistor. 
This issue can be overcome when considering 
nanostructured subunits of graphene: 
0D graphene nanoflake (GNFs) and their 1D counterparts, graphene nanoribbons (GNRs) 
which display a semiconducting gap strongly dependent 
on the system's size.\cite{snookGNF,sonPRL97} 
Particularly interesting is the role of the topology of the edge termination 
of graphene nanostructures in the stabilization of a magnetic state. 
Graphene zigzag (ZZ) edges have a defined chirality and consist of atoms belonging 
to the same triangular sublattice of graphene, 
while in armchair (AC) edges atoms of both sublattices are always paired. 
The unbalance at the ZZ edges is believed to be the origin of magnetism. 
This feature raised the interest for graphene applications 
also in the field of spintronics. 

Recent experimental evidence\cite{wangNL9,ramakrishnaJPCC113,chenSR3,magdaNat514} 
supports the idea that magnetism can be intrinsic in graphene nanostructures, 
and exceptionally high Ne\'{e}l temperatures up to room temperature 
have been reported.\cite{magdaNat514} 
In general, the experimental observation of magnetic states in graphene nanostructures 
remains scarce and controversial, and one of the main difficulties 
in the realization of long-range magnetic structures reside 
in the growth and in the intrinsic irregularity of the sample edges.\cite{chenSR3} 
However, in the last few years, we witnessed important advances in the synthesis 
and in the characterization of graphene nanostructure, 
e.g., at the interface with boron nitride,\cite{drostNL14,drostSR5} 
and the fingerprints of atomically precise edges 
have been uniquely identified in the Raman spectra of GNRs,\cite{verzhbitskiyNL6} 
paving the path toward graphene nanoelectronics. 

From the theoretical point of view there is a substantial agreement 
on the phenomenon of edge magnetism within the framework of density functional 
theory (DFT)\cite{sonNat444,wangNanoLett8,kangJAP112,shengNanoTech21,kabirPRB90} 
and the mean-field approximation of the Hubbard model.\cite{rossierPRL99,wangPRL102} 
In particular, it has been proposed that the magnetic states 
of graphene nanostructures can be exploited for the realization 
of spintronic devices, e.g., spin filters\cite{zhouJPCC116,zouSR5,kangJAP112,shengNanoTech21} 
and logic gates\cite{wangPRL102,zhangSR4} with graphene functional blocks. 
Recent investigation\cite{kabirPRB90,chackoPRB90} suggested 
that the magnetic ordering of the ZZ edges in GNF can be tuned by carrier doping. 
The presence of delocalized charge carriers entails ferromagnetic (FM) correlations,  
giving rise to a complex magnetic phase diagram. 
Indeed, it has been shown that the correlations between spatially separated 
magnetic impurities adsorbed on graphene can be interpreted 
in terms of a Rudermann-Kasuya-Kittel-Yoshida (RKKY) 
exchange interaction mediated by the $\pi$-electrons of the graphene 
substrate.\cite{black-schafferPRB81,szalowskiPRB84,szalowskiPRB90,huPRB84,uchoaPRL106} 

Besides a few relevant 
exceptions,\cite{black-schafferPRB81,feldnerPRB81,feldnerPRL106,duttaSR2,chackoPRB90} 
the role of electronic correlations beyond mean-field theory (MFT) 
in graphene nanostructures remains widely unexplored. 
This calls for better theoretical understanding, 
in particular, on the effect of electronic correlations 
on the magnetic properties and the interplay 
between the charge and spin degrees of freedom in the presence of ZZ edges. 
We address this question in the framework of 
the dynamical mean-field theory (DMFT)\cite{georgesRMP69} 
which is able to describe the interplay between the low-energy coherent excitations, 
arising due to the delocalization of the charge carriers on the lattice, 
and the incoherent high-energy excitations, related to the formation 
of the fluctuating local moment due to the Coulomb interaction. 

The paper in organized as follows. 
In Sec.~\ref{sec:lowe} we discuss the Hubbard model 
as a generic low-energy model for GNFs, 
while in Sec.~\ref{sec:RSDMFT} and \ref{sec:sys-sym} 
we show how DMFT allows to investigate the magnetic properties 
of GNFs in the presence of electronic correlations. 
In Sec.~\ref{sec:results} we present our numerical results. 
We focus on the case of a hexagonal ZGNF and discuss 
the onset of magnetism at half-filling as well as 
the interplay between charge and spin degrees of freedom at finite doping. 
Finally, Sec.~\ref{sec:summary} contains our conclusion and outlook.

\section{Model \& Methods}
\subsection{Low-energy $\pi$-electrons Hamiltonian for GNFs}
\label{sec:lowe} 
In graphene, the in-plane C-C bonds are formed due to a $sp^2$-hybridization 
between carbon $s$, $p_x$, and $p_y$ atomic orbitals, 
while the $p_z$ orbitals are perpendicular to the $sp^2$ bonds 
and bind into $\pi$ orbitals that extend over the plane.   
Hence, in order to describe a GNF we can consider the following Hubbard Hamiltonian 
as an effective low-energy model for the delocalized $\pi$-electrons 
on a finite-size honeycomb lattice with $N$ sites
\begin{equation} \label{eq:lowe} 
{\cal H} = -\sum_{ij\sigma} t_{ij} c^{\dagger}_{i\sigma} c^{\phantom{\dagger}}_{j\sigma} 
         - \mu \sum_{i\sigma} c^{\dagger}_{i\sigma} c^{\phantom{\dagger}}_{i\sigma} 
         +   U \sum_{i} n_{i\uparrow} n_{i\downarrow}. 
\end{equation}
In this notation, the operator $c^{(\dagger)}_{i\sigma}$ annihilates (creates) 
a $\pi$-electron on site $i$ with spin $\sigma$ 
and $n_{i\sigma}=c^{\dagger}_{i\sigma} c^{\phantom{\dagger}}_{i\sigma}$ 
is the corresponding number operator; 
$t_{ij}$ are the tight-binding hopping parameters, 
$\mu$ is the chemical potential, and $U$ denotes the local Coulomb repulsion. 

\begin{figure}[b]
\begin{center}
\includegraphics[width=0.28\textwidth, angle=0]{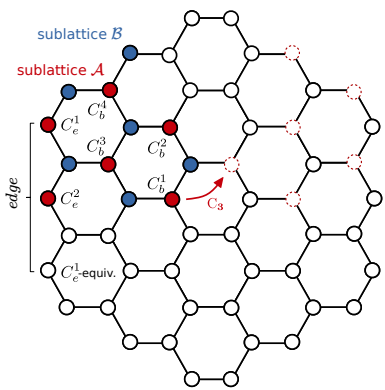}
\caption{(Color online) Schematic representation of the hexagonal ZGNF considered.  
The $N_{\rm ineq}=6$ inequivalent C atoms are distinguished into 
bulk- ($C_b^{1,...,4}$) and edge-atoms ($C_e^{1,2}$) 
for each sublattice ${\cal A}$ (red/dark grey) and sublattice ${\cal B}$ (cyan/light grey). 
The rotational symmetry $C_3$ sends each inequivalent atoms 
to their equivalent atoms of the same sublattice, indicated with dashed circles. }
\label{fig:3nhexaZGNF}
\end{center}
\end{figure}

The information on the spatial arrangement of the C atoms 
in the nanostructure is contained in the real-space hopping matrix, 
including also the topology of the edges (either ZZ or AC). 
Here, we restrict ourselves to consider the case  
of the hexagonal ZGNF shown in Fig.~\ref{fig:3nhexaZGNF},   
while spatial symmetries are discussed in detail in Sec.~\ref{sec:sys-sym}. 
We assume a homogeneous hopping $t_{ij}=t$, 
where the nearest-neighbor (NN) hopping amplitude $t\equiv1$ 
sets the energy scale of the system, 
and we neglect hopping processes beyond NN. 
Recently, Kretinin {\it et al.}\cite{kretininPRB88} 
experimentally estimated the value of the next-NN hopping parameter in graphene 
to be $t'/t\approx0.1$. While the presence of $t'$ have important consequences 
as breaking the particle-hole symmetry of the Hamiltonian, 
it was concluded that the asymmetry leads to relatively 
weak effects in the optical, as well as in the electronic, 
and presumably spin transport properties of monolayer graphene.\cite{kretininPRB88} 
A configuration with spatially uniform hopping parameters 
is representative of the case where all the dangling C-C bonds at the ZZ edges 
are passivated, e.g., with hydrogen atoms. 
Within DFT is was shown that 
passivation quenches significantly the edge magnetic moments, 
while the lack of passivation changes the $sp^2$-hybridization between C atoms 
and induces sizable lattice distortions, 
mostly at the edges but also in the bulk.\cite{kabirPRB90} 
A full structure relaxation allows to derive the DFT tight-binding parameters, 
i.e., the hopping amplitudes and the local crystal fields of the distorted structure. 
However, according to the numerical results, 
both neutral and hole-doped case (and in contrast to the electron-doped one) 
the ZGNF does not display sizable lattice distortions, 
and the doped charges are distributed symmetrically over the edges.\cite{kabirPRB90} 
In the following analysis we focus on hole-doped ZGNFs 
in order to study the interplay between charge and spin degrees of freedom 
in the stabilization of different magnetic phases. 
Hence, we disregard the effects of lattice distortions 
as we do not expect any qualitative change in the results obtained. 
Finally, we consider a local Coulomb interaction $U$ 
between the delocalized $\pi$-electrons. 
Recently, both local and non-local Coulomb repulsion terms have been estimated 
to be sizable in graphene,\cite{wehlingPRL106} 
justifying the necessity to treat graphene beyond the tight-binding 
or mean-field approximation. 
Indeed, electronic correlations, as well as the interplay 
between local and non-local repulsive interactions 
are expected to play an important role in the stabilization 
of different magnetic orders in graphene nanostructures.\cite{chackoPRB90} 
In the following we shall focus on the dynamical correlation effects 
driven by the local repulsion $U$ within the framework of DMFT.\cite{georgesRMP69}  
Unless specified otherwise, we choose a typical value of $U=3.75t$, 
in line with recent estimates for graphene.\cite{wehlingPRL106,schuelerPRL111} 
Non-local interaction could be taken into account within DMFT 
by including in Hamiltonian (\ref{eq:lowe}) a mean-field term 
$V_{ij} \sum_{\sigma} n_{i\sigma} 
                     (\langle n_{j\uparrow}\rangle + \langle n_{j\downarrow} \rangle)$.  
However, the presence of non-local repulsion favors charge modulation on the lattice 
and possibly leads to the proliferation of ordered states, 
which makes this extension beyond the scope of the present work.

\subsection{Real-space dynamical mean-field theory with magnetic symmetry breaking}
\label{sec:RSDMFT}
DMFT is a well established theoretical tool that allows to take into account 
local electronic correlations non-perturbatively. 
Numerous extensions of DMFT have also been proposed 
in which the self-energy is local albeit site-dependent, allowing to deal with 
inhomogeneous\cite{potthoffPRB59,florensPRL99,snoekNJP10,titvinidzePRB86} and 
nanoscopic\cite{valliPRL104,jacobPRB82,valliPRB86,dasPRL107,valliPRB91,valliPRB92} 
systems, where in general the translational symmetry is broken 
along one or more directions in space. 
Non-local electronic correlations beyond mean-field are 
in general expected to be important in low-dimensional system. 
However, by means of comparative studies\cite{valliPRB91} 
with diagrammatic\cite{toschiPRB75} extensions of DMFT 
built on the local two-particle vertex function\cite{rohringerPRB86} 
it has been demonstrated how a reasonable insights 
of electronic and transport properties 
of correlated nanostructures\cite{valliPRL104,valliPRB86,valliPRB91} 
are already gained at the DMFT level. 
In the following we briefly recall how DMFT is implemented for 
an inhomogeneous finite system, and we discuss 
how to handle magnetic phases within this framework.  
In the case of a finite system, one can map each site $i=1,...,N$ 
of the original many-body problem onto an auxiliary Anderson impurity model (AIM) 
embedded in a self-consistent bath determined by the rest of the system. 
The auxiliary AIM for the $i$-th site is defined by the 
spin-dependent local dynamical Weiss field ${\cal G}_{0i\sigma}(\omega)$ 
in terms of the local element of the real-space Green's function $G_{ij\sigma}(\omega)$ 
of the whole system and the local self-energy $\Sigma_{i\sigma}(\omega)$ as 
\begin{equation} \label{eq:weissRDMFT}
{\cal G}^{-1}_{0i\sigma}(\omega) = G_{ii\sigma}^{-1}(\omega) + \Sigma_{i\sigma}(\omega).
\end{equation}
In general, the local problems defined by ${\cal G}^{-1}_{0i\sigma}(\omega)$ 
are inequivalent, and each of them can be solved numerically 
yielding a local dynamical self-energy $\Sigma_i(\omega)$ 
which carries a spatial dependence on the site index $i$. 
However, one can exploit any spatial symmetry of the original system 
and reduce the numerical effort by solving, eventually, 
only a subset of $N_{\rm ineq} \leq N$ inequivalent local problems. 
This reduced the complexity of the problem 
from exponential in $N$ to linear in $N_{\rm ineq}$. 
The knowledge of all (inequivalent) $\Sigma_i(\omega)$ allows to compute 
the Green's function of the whole system from the real-space Dyson equation
\begin{equation}
 G^{-1}_{ij	\sigma}(\omega) = (\omega+\mu)\delta_{ij} - t_{ij} 
                           - \Sigma_{i\sigma}(\omega)\delta_{ij}, 
\end{equation}
where the self-energy matrix only contains the local, site-dependent elements.  
Non-local correlations between different sites are neglected. 
From the Green's function one can define a new set of auxiliary AIMs 
and iterate the above process self-consistently until convergence. 

In order to study the emergence of magnetism, we lift 
the local $SU(2)$ spin rotational symmetry of the auxiliary AIM, 
and allow the impurity solver to access solutions with 
a finite on-site magnetization  
$\langle S^z_i\rangle = \langle n_{i\uparrow} - n_{i\downarrow} \rangle$ 
(here we only consider solutions with magnetization in the $z$ direction). 
This is done locally for each site $i$ 
using a symmetry-broken Weiss field ${\cal G}_{0i\sigma}$ 
as initial input for DMFT. 
In this respect, the separate treatment of the different spin directions 
is the only essential modification of the general self-consistent scheme 
of real-space DMFT, and in particular the self-consistent equations,  
whatever is the magnetic phase to be investigated, 
as opposed to the standard implementation of symmetry-broken 
solutions within DMFT.\cite{georgesRMP69} 
The landscape of the possible magnetic phases that can be explored 
within this approach depends on the set of spatial symmetries 
enforced in the calculation 
and on the specific choice of the initial symmetry-breaking.

\subsection{Spatial symmetries and magnetic phases of hexagonal ZGNFs}
\label{sec:sys-sym}
In the following we discuss in detail the spatial symmetries of hexagonal ZGNFs 
which we enforce in order to investigate a landscape of possible magnetic configurations 
within the self-consistent DMFT calculations. 
We consider the hexagonal ZGNF shown in Fig.~\ref{fig:3nhexaZGNF}, 
which consists of a bipartite honeycomb lattice with $N=54$ C atoms. 
Exploiting both the rotational symmetry of the $C_{3v}$ point-group 
and the sublattice symmetry, one can identify $N_{\rm ineq}=6$ inequivalent C atoms 
all belonging to the same triangular sublattice (e.g., sublattice ${\cal A}$). 
The inequivalent C atoms can be further distinguished into 
bulk-atoms (denoted as $C_b^{1-4}$), which have three in-plane $sp^2$ C-C bonds, 
and edge-atoms (denoted as $C_e^{1,2}$), which have two C-C bonds 
and one dangling/passivated bond. 
A ZZ edge of the ZGNF consists of $N_{\rm edge}=3$ C atoms, 
i.e., two (equivalent) $C_e^{1}$ atoms and a $C_e^{2}$ atom, 
all belonging to the same sublattice. 
Neighboring edges consist of C atoms belonging to different sublattices, 
and are always connected by an AC bond between $C_e^{1}$ atoms. 
Lifting the local $SU(2)$ spin rotational symmetry 
would be enough to study, e.g., (inhomogeneous) FM. 
However, as the Hubbard model on a bipartite lattice has a natural tendency 
toward a N\'{e}el AF state (close to half-filling), 
a natural choice would be to enforce each kind of inequivalent atom  
to have opposite magnetization on different sublattices. 
However this assumption would not allow other magnetic configurations, 
and in particular FM. 
A more general description of the magnetic phases requires 
instead to raise the number of inequivalent atoms in the system. 
Here we choose to lift the sublattice symmetry, 
i.e., treat each inequivalent atom 
and its counterpart in the other sublattice independently, 
thus raising $N_{\rm ineq}: 6 \rightarrow 12$. 
This choice allows us to stabilize either an AF or a FM state 
and describe the competition between the two short-range magnetic orders 
emerging from the interplay 
between charge and spin degrees of freedom at finite doping.

\section{Results and Discussion} 
\label{sec:results}
In the following sections we discuss the onset of AF insulating state  
of a ZGNF at half-filling. 
We also show that at finite doping there exist another magnetic state 
underneath the AF one, in which the magnetic moments at the ZZ edges are aligned FM. 
Such a state is unstable with respect to temperature fluctuations.  
We discuss its possible origin of the magnetic transition 
analyzing the effective magnetic exchange interaction mediated by the charge carriers.  

\subsection{ZGNF at half-filling}
\label{sec:hf}
At half-filling, which corresponds to an average occupation 
of $\langle n\rangle=1$ electrons/site, 
and for passivated edges (i.e., in the case of homogeneous hopping parameters), 
Hamiltonian (\ref{eq:lowe}) is particle-hole symmetric 
and the density of states of the ZGNF is symmetric 
with respect to the chemical potential. 
Due to the discreteness of the spectrum, 
even in the absence of Coulomb interaction the system is semiconducting, 
with a charge gap $\Delta_0 \approx 0.7t$. 
The value of $\Delta_0$ depends on the system's size and shape, 
and in particular it has been shown 
both experimentally\cite{ritterNM8} and theoretically\cite{huJCP141} 
that it decreases as $1/L$ with the linear size $L$ of the GNF 
and vanish toward the semimetallic limit 
realized in bulk graphene. 

\begin{figure}[b]
\includegraphics[width=0.45\textwidth, angle=0]{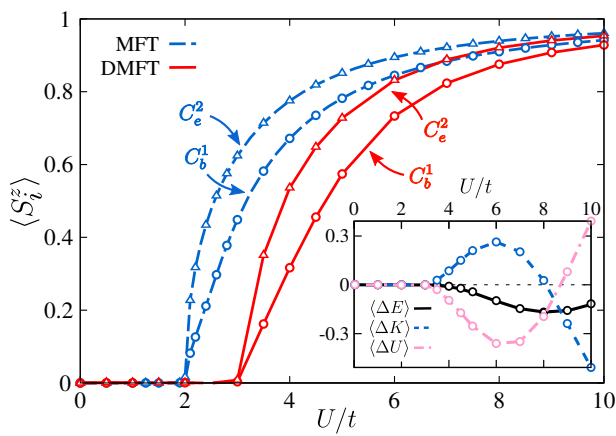}
\caption{(Color online) Onset of the AF order at $T=0$ 
within static MFT (cyan/light grey) and DMFT (red/dark grey).  
The local ordered magnetic moment 
$\langle S^z_i \rangle = \langle n_{i\uparrow}-n_{i\downarrow} \rangle$ 
for representative bulk $C_b^{1}$ and edge $C_e^{2}$ atoms 
displays a clear bulk-edge dichotomy. 
[Inset] DMFT energy balance between the fully-compensated AF and the PM phases. 
The total energy difference  
$\langle \Delta E \rangle = \langle H \rangle_{\rm AF} - \langle H \rangle_{\rm PM}$ 
is separated into kinetic $\langle \Delta K \rangle$ 
and potential energy $\langle \Delta U \rangle$ contributions. }
\label{fig:onsetAF}
\end{figure} 

We characterize the onset of the AF state at half-filling and at $T=0$ 
by comparing the results obtained within static mean-field theory (MFT) and DMFT. 
In the main panel of Fig.~\ref{fig:onsetAF} we show the local magnetic moment 
$\langle S^z_i \rangle = \langle n_{i\uparrow}-n_{i\downarrow} \rangle$ 
for representative atoms in the ZGNF, 
i.e., for bulk $C_b^{1}$ and edge $C_e^{2}$ atoms, 
as a function of the local interaction $U/t$. 
It is interesting to notice that the onset of AF happens 
at a finite value of $U/t$ and simultaneously for all inequivalent atoms, 
even though the size of the ordered moment of $C_b^{1-4}$ 
(which are all similar, yet not identical) 
is different from that of $C_e^{1}$ and $C_e^{2}$. 
Hence, we observe a clear \emph{dichotomy} between bulk and edge atoms, 
even in the passivated case, which persists also when increasing the interaction. 
The resulting magnetic state is a fully-compensated AF state 
but it is different from the conventional N\'{e}el state 
due to the inhomogeneous spatial distribution of the magnetic moments. 
Both static MFT and DMFT show the qualitative trend discussed above. 
Unsurprisingly, dynamical quantum effects suppress the AF phase, 
pushing the onset interaction toward the strong coupling regime, 
i.e., from the value $U_{\rm AF}\approx 2t$ obtained within static MFT 
to the value $U_{\rm AF}\approx 3t$ obtained within DMFT. 
Feldner {\it et al.}\cite{feldnerPRB81} have shown that static MFT 
overestimates both the size of the local magnetic moment and the spectral gap 
of half-filled ZGNFs with respect to 
exact diagonalization and Quantum Monte Carlo simulations. 
It is also interesting to notice that the relative difference in size 
between the magnetic moment of bulk and edge atoms 
is enhanced within DMFT with respect to static MFT. 
In fact, the spatial variation of the ordered local moment 
$\langle S^z_i \rangle$ can be traced back  
to a preformed (disordered) local moment in the paramagnetic (PM) state 
$\langle p_i \rangle = \langle(S^z_i)^2 \rangle 
                               = \langle n_{i\uparrow}+n_{i\downarrow} \rangle 
                                - 2\langle n_{i\uparrow} n_{i\downarrow} \rangle$, 
which already displays the bulk/edge dichotomy. 
In the magnetic state the value of $\langle p_i \rangle$ increases 
due to the decrease of the double occupation 
$\langle n_{i\uparrow} n_{i\downarrow} \rangle$, 
as each sites stays locally half-filled, i.e., $\langle n_i \rangle =1$.  
In the inset of Fig.~\ref{fig:onsetAF} we show the DMFT energy balance 
as a function of $U/t$, where the internal energy difference 
between the fully-compensated AF and the paramagetic phases, 
i.e., $\langle \Delta E \rangle = \langle H \rangle_{\rm AF} - \langle H \rangle_{\rm PM}$ 
is separated into kinetic $\langle \Delta K \rangle$ 
and potential energy $\langle \Delta U \rangle$ contributions. 
For realistic values of the interaction parameter in graphene, 
the AF state is stabilized by a gain of potential energy $\langle \Delta U \rangle<0$ 
corresponding to the reduction of the double occupation upon ordering. 
The above scenario mirrors the well-known 
DMFT picture\cite{toschiPRB72,tarantoPRB85,rohringer1604.08748,tagliavini1604.08882} 
of the AF transition in the bulk Hubbard model, 
with its crossover from weak-to-strong coupling physics 
at values of $U$ of the order of the bandwidth. 
This consideration would put any realistic value of the interaction 
in ZGNF definitely on the weak-coupling (Slater) side. 
The main difference here is the that the AF phase 
is not stabilized at arbitrary weak coupling, 
but it requires a finite onset interaction $U_{\rm AF}$ 
due to the semiconducting nature of the ZGNF at half-filling. 
Let us note that the value of the onset interaction depends on 
the size of the (correlated) spectral gap $\Delta$ in the PM phase, 
which shrinks with the linear size $L$ of the ZGNF.\cite{ritterNM8,huJCP141} 
Thus ZGNF with increasing size are expected to become magnetic 
at weaker interaction, while for small ZGNF the onset interaction $U_{\rm AF}$ 
is dominated by finite-size effects. 
Further increasing the size the semimetallic nature of graphene plays a role. 
In fact, despite $\Delta \rightarrow 0$, 
the lack of perfect nesting on the honeycomb lattice 
and the zero density of states at the Dirac point keep the onset interaction finite. 
Theoretical estimates of the onset interaction 
range from $U_{\rm AF}\approx3.8t$ to $U_{\rm AF} \approx 4.5t$ 
with different numerical techniques,\cite{sorellaEPL19,sorellaSR2,taoSR4,aryaPRB92} 
which seems to be in agreement 
with the experimental absence of AF in graphene monolayers.

\begin{figure}[t]
\includegraphics[width=0.37\textwidth, angle=0]{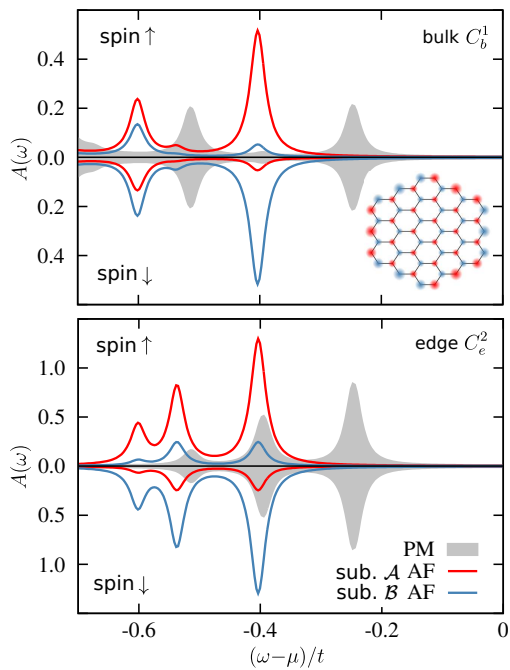}
\caption{(Color online) Local spin-resolved DMFT spectral function $A(\omega)$ 
for bulk $C_b^{1}$ (upper panel) and edge $C_e^{2}$ (lower panel) atoms 
at $U/t=3.75$, $\langle n \rangle=1$, and $T=0$. 
In the non-magnetic calculation the ZGNF is a semiconductor (grey shaded area) 
while the magnetic calculation yields a fully-compensated AF insulating state, 
with opposite spin polarization in sublattice ${\cal A}$ (red/dark grey solid line) 
and ${\cal B}$ (cyan/light grey solid line). 
The spatial distribution of the magnetic moments on the ZGNF 
is represented in the inset,  where the color and the radius 
of the circles indicate the sign and the magnitude of $\langle S^z_i \rangle$. }
\label{fig:awPM2AFn1T0}
\end{figure} 

\begin{figure}[t]
\includegraphics[width=0.45\textwidth, angle=0]{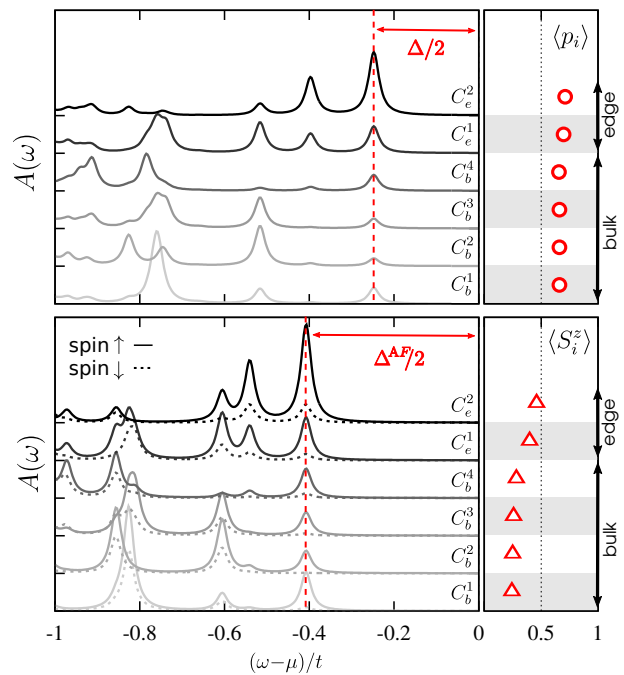}
\caption{(Color online) Local spin-resolved DMFT spectral function $A(\omega)$ 
for all inequivalent C atoms in the PM (upper panel) and AF (lower panel) state 
at $U/t=3.75$, $\langle n \rangle=1$, and $T=0$.  
The electronic coherence at low-energy determines 
homogeneous spectral gaps $\Delta$ and $\Delta^{\rm AF}$ (vertical dashed lines), 
despite the disordered $\langle p_i \rangle$  
and ordered $\langle S^z_i \rangle$ local moments (open symbols in the side panels) 
display a clear bulk-edge dichotomy. }
\label{fig:aw-ineq_leh}
\end{figure} 

\begin{figure*}[t]
\includegraphics[width=0.90\textwidth, angle=0]{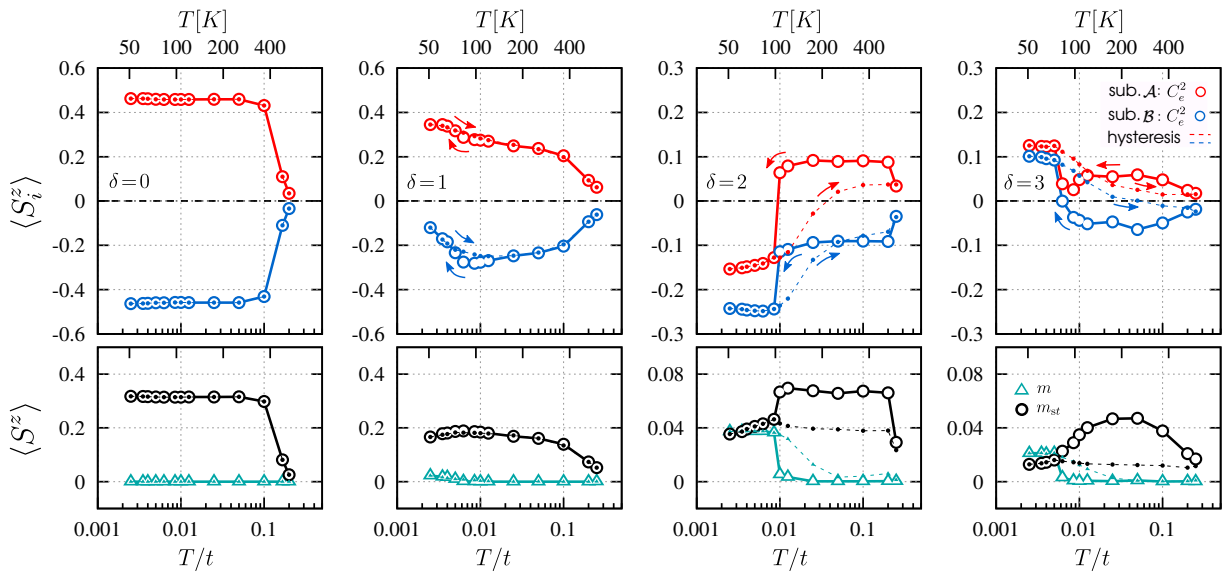}
\caption{(Color online) Evolution of the magnetic moments in the ZGNF 
as a function of temperature $T/t$ at $U/t=3.75$ and different values of doping $\delta$. 
As a reference, the temperature scale in K is obtained considering 
a realistic value $t=2.7$~eV for the hopping parameter in graphene. 
[Upper panels] Local magnetic moment $\langle S^z_i \rangle$ 
for the edge $C_e^{2}$ atoms in sublattice ${\cal A}$ (red/dark grey symbols) 
and sublattice ${\cal B}$ (cyan/light grey symbols). 
[Lower panels] Absolute value of the net magnetic moment $m$ (jade/light grey triangles) 
and the staggered magnetization $m_{st}$ (black circles) per atom in the ZGNF. 
The dashed line with filled symbols show the hysteretic behavior of the magnetization. }
\label{fig:mvsT_doping}
\end{figure*} 

It is interesting to discuss in detail the change in the low-energy 
spectral properties of the half-filled ZGNF across the magnetic transition. 
In Fig.~\ref{fig:awPM2AFn1T0} we show the occupied portion 
of the local spin-resolved spectral function $A(\omega<0)$ 
for representative bulk $C_b^{1}$ and edge $C_e^{2}$ atoms. 
Due to the particle-hole symmetry, the spectral function for spin $\sigma$ 
fulfills the relation $A_{\sigma}(\omega>0)=A_{\overline{\sigma}}(-\omega)$. 
We consider a local interaction $U=3.75t$, which at $T=0$ lies 
above but close to the DMFT onset value $U_{\rm AF}$. 
In the absence of magnetism, the ZGNF in Fig.~\ref{fig:awPM2AFn1T0} 
is semiconducting, and the local spectral function (grey shaded area) 
displays a spectral gap $\Delta \approx 0.5t$, 
where the gap is defined as the distance between the lowest energy peaks 
around the Fermi level. 
we notice that the spectral gap $\Delta$ 
is substantially \emph{reduced} by local electronic correlations 
with respect to the tight-binding value $\Delta_0 \approx 0.7t$, 
as expected in view of similar observations (within DMFT) 
in the insulating state of both bulk crystal\cite{sentefPRB80} 
and molecules.\cite{valliPRB91} 
In the fully-compensated AF phase we plot $A(\omega)$ for atoms in both sublattices, 
to show that the spin-$\uparrow$ and spin-$\downarrow$ spectral functions 
are inverted between the sublattices as a consequence of the particle-hole symmetry, 
which is fulfilled at half-filling. 
It is important to notice that both the gap in the PM state $\Delta$ 
and the AF gap $\Delta^{AF}$ do not display 
any spatial dependence over the ZGNF,  
despite the local magnetic moment of bulk and edge C atoms being different. 
This is shown in Fig.~\ref{fig:aw-ineq_leh}, 
where we plot the site-resolved spectral functions 
for all inequivalent atoms of sublattice $\cal A$ 
in the PM state (upper panel) and for both spin-$\uparrow$ and spin-$\downarrow$ 
in the fully-compensated AF state (lower panel). 
The vertical dashed line indicates the position of the lowest-energy peak 
for both spin polarizations. 
The side panels show also the corresponding 
disordered local moment $\langle p_i \rangle$ 
and magnetic moment $\langle S_i^z \rangle$. 
The homogeneity of the spectral gap is the fingerprint 
of the separation between low-energy delocalized excitations 
and high-energy localized states in strongly correlated systems. 
The high-energy properties follow the inhomogeneity 
dictated by the geometry or the single-particle potential 
to minimize the potential energy, while the low-energy properties, 
and in particular the spectral gap, 
are more homogeneous as they are associated 
to a delocalized behavior which lowers the kinetic energy.\cite{amaricciPRA89}
In this respect, we can conclude that, 
for the linear size of the ZGNF that we have considered here,  
one can observe both the finite-size effects, which result 
in the physics being dominated by the ZZ edges, 
but also electronic features that would be expected in the bulk 
of an infinitely extended system. 

Extending the analysis to finite temperature, 
as also discussed in detail in the following section, 
we find the fully-compensated AF state at half-filling to be stable 
up to room-$T$ (see first panel of Fig.~\ref{fig:mvsT_doping}), 
in agreement with recent experimental evidence in ZGNRs.\cite{magdaNat514} 
Moreover, the properties of the ordered state, and in particular 
the local magnetic moments $\langle S^z_i \rangle$ 
display a very weak dependence on $T$ in the whole range of $T$ explored. 
 

\subsection{Competing magnetic orders upon doping}
\label{sec:dope}
In the following we explore the interplay between charge and spin degrees of freedom 
upon hole doping.\cite{note:doping} 
Charge carries can be introduced in the ZGNF, e.g., by using a gate electrode 
or by chemical substitution with carboxyl (COOH) or hydroxyl (OH) groups, 
which should not disrupt the $sp^2$ hybridization at the edges.\cite{moonRSCA6}
The most interesting result is that the fully-compensated AF state 
is unstable upon doping, due to the emergence 
of ferromagnetic (FM) correlations between spins at the ZZ edges. 
We show that at finite doping and below a critical temperature $T_c$ 
it is energetically favorable for the local magnetic moment 
$\langle S^z_i \rangle$ of the $C_e^{2}$ atoms 
to be aligned FM both within the same edge and between neighboring edges, 
while bulk C atoms tend to maintain an AF pattern. 
The resulting magnetic state is characterized 
by a uncompensated net magnetic moment \emph{and} a finite staggered magnetization. 
In the following we denote it as ferrimagnetic (FI) state, although 
we stress that the ZGNF does not display a proper FI ordered. 
A similar behavior upon doping was recently observed for the same ZGNF 
within DFT calculations at $T=0$.\cite{kabirPRB90}

\begin{figure}[t]
\includegraphics[width=0.37\textwidth, angle=0]{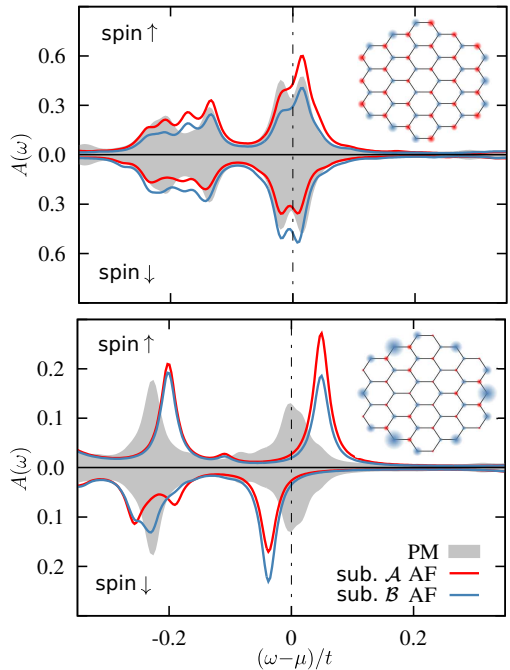}
\caption{(Color online) Local spin-resolved DMFT spectral function $A(\omega)$ 
for the edge $C_e^{2}$ atoms at $U/t=3.75$ and $\delta=2$. 
The data are shown for $T \approx 160$~K (upper panel) and $T \approx 80$~K (lower panel) 
which are representative of the high-$T$ fully-compensated AF state 
and the low-$T$ FI state, respectively. 
The color coding reads as in Fig.~\ref{fig:awPM2AFn1T0} 
for both the main panels and the inset. }
\label{fig:awPM2AFn0963Tf}
\end{figure} 
\begin{figure}[t]
\includegraphics[width=0.37\textwidth, angle=0]{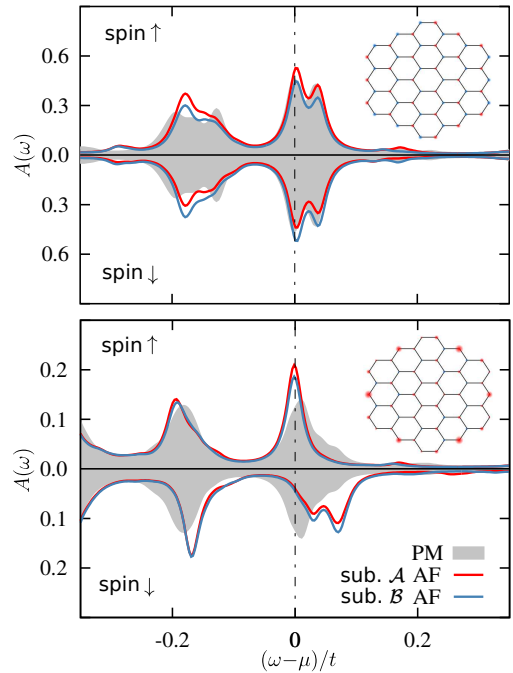}
\caption{(Color online) As in Fig.~\ref{fig:awPM2AFn0963Tf} but for $\delta=3$, 
showing a FM alignment of the $C_e^{2}$ magnetic moments below $T_c$. }
\label{fig:awPM2AFn0945Tf}
\end{figure}

The upper panels of Fig.~\ref{fig:mvsT_doping} show 
the temperature evolution of the local magnetic moment $\langle S^z_i \rangle$ 
for the edge $C_e^{2}$ atoms of sublattice ${\cal A}$ and ${\cal B}$. 
We provide also reference values of $T$ in K, obtained with a typical value 
$t \approx 2.7$~eV for the hopping integral in bulk graphene. 
The doping is denoted by $\delta$, the integer number of holes in the ZGNF, 
so that nominal filling of the ZGNF is $n=(N-\delta)/N$.  
For all the cases we considered, the magnetic state in the high-$T$  
is a the fully-compensated AF state. 
The orientation of the magnetic moment is opposite 
for atoms in different sublattices, 
giving rise, globally, to a staggered magnetization 
$m_{st} = \frac{1}{N}\sum_{i=1}^{N} \langle S^z_i \rangle^{i}$. 
Away from half-filling and below a doping-dependent temperature $T_c(\delta)$ 
the system also develops a finite net magnetic moment 
$m = \frac{1}{N}\sum_{i=1}^{N} \langle S^z_i \rangle$ 
(uniform magnetization) which coexists with a finite $m_{st}$ 
giving rise to a FI state. 
The results for the absolute value of the magnetic moments $m$ and $m_{st}$ are shown 
in the lower panels of Fig.~\ref{fig:mvsT_doping} for each $\delta$. 
At half-filling the fully-compensated AF state is characterized 
by a N\'{e}el temperature $T_N \approx T_{\rm room}$ 
and does not display any tendency toward the FI state 
down to the lowest $T$ explored. 
While the mean-field character of the spatial fluctuations within DMFT 
is known to overestimate the ordering temperature,\cite{rohringerPRL107,hirschmeierPRB92}
the observation of a sizable $T_N$ for edge magnetism 
is in agreement with recent experimental evidence in ZGNR.\cite{magdaNat514} 
The annealing procedure at $\delta \neq 0$ shows that 
the ZGNF is driven away from a fully-compensated AF state upon lowering $T$  
by breaking the spin inversion symmetry 
between the ${\cal A}$ and ${\cal B}$ sublattices. 
In the case of one hole, i.e., $\delta=1$ 
(second panel from the left in Fig.~\ref{fig:mvsT_doping}), 
the local magnetic moments $\langle S^z_i \rangle$ for the $C_e^{2}$ atoms 
increases, as to be expected, upon lowering $T$ 
until the sublattice symmetry is broken at $T_c/t \approx 0.008$ ($T_c \approx 100$~K). 
Below $T_c$, the staggered magnetization $m_{st}$ decreases 
and the ZGNF develops a net magnetic moment $m \neq 0$. 
The major contribution to $m$ (and to the decrease of $m_{st}$) is given by 
the asymmetry that develops between the magnetic moments 
of the $C_e^{2}$ atoms in the two sublattices. 
Let us stress once again that the magnetic transition 
happens \emph{spontaneously} upon annealing, 
as the symmetries of the ZGNF enforced in the numerical calculation 
allow both the AF and the FM solutions, as well as the coexistence of the two orders. 
The situation is substantially different at higher hole concentrations. 
At $\delta=2$ (third panel in Fig.~\ref{fig:mvsT_doping}) 
the ZGNF displays a sudden change in the magnetic configuration 
at $T_c/t \approx0.02$ ($T_c \approx 140$~K) 
from a high-$T$ fully-compensated AF state to a low-$T$ FI state. 
In the FI state all the inequivalent $C_e^{2}$ spins are aligned FM, 
although they are not equal in size due to the breaking 
of the sublattice symmetry. 
The trend of the data at low $T$ suggests that the sublattice asymmetry 
could possibly disappear in the limit $T\rightarrow 0$. 
Away from the transition, the properties of the ZGNF 
are nearly independent on $T$ up to a sharp drop above $T_{\rm room}$. 
A similar behavior is found at $\delta=3$ 
(fourth panel in Fig.~\ref{fig:mvsT_doping}). 
We can estimate the transition temperature to $T_c \approx 100$~K, 
which results to be lower than the one for $\delta=2$. 
Interestingly, the effect in the bulk atoms in the ZGNF is weaker, 
although a clear discontinuity in the $T$ dependence 
of the local magnetic moment can be observed for $\delta=2$ and $\delta=3$ (not shown). 

At $\delta \neq 0$ the magnetization also displays a hysteretic behavior. 
The hysteresis is evident especially for $\delta=2$ and $\delta=3$, 
where it extends over a wide range of $T$ 
and indicates a coexistence of short-range AF and FM orders. 
We notice that both the local magnetic moments $\langle S^z_i \rangle$ 
decrease upon doping. While this is expected within DMFT, 
it is instead absent in reference $T=0$ DFT calculations, 
where the spin density at the edges is the same 
in both the AF and the FI states.\cite{kabirPRB90} 

It is interesting to relate the changes in the local magnetic moments 
at the ZZ edges across the AF-to-FI transition, with the changes 
in the low-energy excitation in the spectral functions. 
To this end, we focus on the edge $C_e^{2}$ atoms of sublattices ${\cal A}$ and ${\cal B}$. 
The corresponding local spin-resolved spectral functions $A(\omega)$ 
in the PM and the magnetic (AF or FI) states 
are shown in Figs.~\ref{fig:awPM2AFn0963Tf} and \ref{fig:awPM2AFn0945Tf} 
at $\delta=2$ and $\delta=3$, respectively. 
A common feature of doped ZGNFs is the metallic character of the spectrum 
due to a redistribution of spectral weight in the site-resolved $A(\omega)$ 
with respect to half-filling (compare with Fig.~\ref{fig:awPM2AFn1T0}).   
We observe a resonance at the Fermi level, associated with  
the delocalization of the electrodoped charge carriers on the lattice. 
The low-energy coherent excitations at $\delta \neq 0$ 
coexist with the incoherent high-energy excitations (Hubbard bands) 
related to the formation of the fluctuating local moment 
due to the Coulomb interaction. 
In the magnetic state above $T_c(\delta)$ the analysis of the spectral functions 
in (the upper panels of) Figs.~\ref{fig:awPM2AFn0963Tf} and \ref{fig:awPM2AFn0945Tf}   
clearly indicates the AF alignment of the $C_e^{2}$ atoms in the different sublattices. 
Below $T_c$ (lower panels) a redistribution of low-energy spectral weight 
splits the spin-$\uparrow$ and spin-$\downarrow$ spectral functions 
with respect to the Fermi level, showing a tendency toward an insulating state. 
The splitting is the same for sublattice ${\cal A}$ and ${\cal B}$. 
As a consequence the local magnetic moments of the edge $C_e^{2}$ atoms 
in the two sublattices are aligned FM. 
For $\delta=2$ the sublattice asymmetry is evident while for $\delta=3$ is minimal. 
The spatial distribution of the magnetic moments above and below $T_c$, 
and the corresponding change of the magnetic pattern for both values of $\delta$ 
are shown in the respective insets.

\begin{figure}
\includegraphics[width=0.48\textwidth, angle=0]{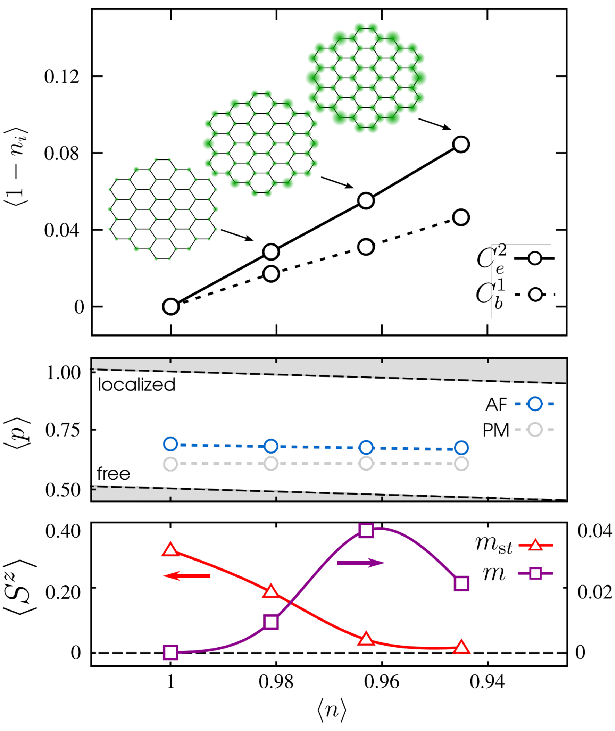}
\caption{(Color online) Distribution and influence of the doped holes 
as a function of the average electron density in the ZGNF $\langle n \rangle$ 
at $U/t=3.75$ and $T \approx 80$~K. 
[Upper panel] Local hole density $\langle 1 - n_i \rangle$ 
for the bulk $C_b^{1}$ and the edge $C_e^{2}$ atoms in the PM state. 
The spatial distribution of the holes in the ZGNF is shown in the insets. 
[Middle panel] The average fluctuating local moment 
$\langle p \rangle$. 
The dashed lines separating the shaded area 
correspond to the free and fully localized limits (see text for the details). 
[Lower panel] Absolute value of the net magnetic moment $m$ 
and staggered magnetization $m_{st}$. }
\label{fig:npmvsn}
\end{figure}

In order to understand the nature of the magnetic correlations at the ZZ edges 
we analyze the spatial distributions of the holes upon doping. 
In the upper panel of Fig.~\ref{fig:npmvsn} we show the local hole density 
for the bulk $C_b^{1}$ and edge $C_e^{2}$ atoms as a function 
of the average electron density $\langle n \rangle$ in the ZGNF. 
The holes are found to be localized mostly at the ZZ edges,  
and the ratio between the hole concentration 
at the edge and in the bulk increases with doping. 
However, the average hole concentration on the lattice obtained within DMFT 
is less heterogeneous than in reference DFT calculations\cite{kabirPRB90} 
due to the effects of the hole-hole repulsion at the edges. 
The hole concentration in Fig.~\ref{fig:npmvsn} 
is shown for $T/t=0.005$ ($T \approx 80$~K), 
which is below $T_c(\delta)$ for all $\delta \neq 0$, 
but the spatial distribution of the holes on the lattice 
is very weakly dependent on $T$ (not shown). 
We also find that there is no sizable redistribution of the holes on the lattice 
between the PM and the magnetically ordered states, except for a slight asymmetry 
due to the sublattice symmetry breaking in the FI state. 
In the other two panels of Fig.~\ref{fig:npmvsn}, 
we show the evolution with doping of the average fluctuating local moment 
$\langle p \rangle = \frac{1}{N} \sum_{i} \langle p_i \rangle$ 
in relation with the magnetization $m$ and $m_{st}$ in the ordered state. 
Upon ordering magnetically, the value of $\langle p \rangle$ 
increases due to the reduction of the double occupations.  
At half-filling this corresponds to the gain of potential energy 
shown in the inset of Fig.~\ref{fig:onsetAF}. 
We note that $\langle p \rangle$ is weakly dependent on doping. 
If we compare it with the value $\langle p \rangle$ 
in the uncorrelated (free) and the fully localized cases, 
where the local double occupations are 
$\langle n_{i\uparrow} n_{i\downarrow} \rangle = 0.25$ and
$\langle n_{i\uparrow} n_{i\downarrow} \rangle = 0$, respectively. 
we can conclude that $\langle p \rangle$ gets closer to the localized limit upon doping. 
At the same time the magnetization $m_{st}$ is strongly suppressed 
in favor of a uncompensated magnetic moment $m$ 
as FM correlations tend to align the magnetic moment at the ZZ edges. 
The net magnetic moment as a function of doping $m(\delta)$ 
displays a dome shape, peaked at an optimal value of $\delta \approx 2$, 
which develops upon lowering $T$. 
Within the usual DMFT picture, the presence of a sizable 
\emph{preformed} local moment while the magnetic order is capped 
by a lower coherence energy scale, would indicate the realization 
of a strong-coupling scenario.\cite{toschiPRB72,tarantoPRB85} 
This suggests a crossover from weak- to strong-coupling magnetism 
in ZGNF upon doping away from half-filling. 
Based on these observations we can argue that 
the delocalized holes mediate an effective magnetic exchange interaction, 
which is dynamically generated between the magnetic moments localized at the edges.  
This mechanism ultimately leads to the change 
of the magnetic structure in the doped ZGNF. 
This highly non-trivial physics can be indeed captured by DMFT 
because it is able to describe both the coherent and the incoherent excitations 
as well as their interplay. 
Evidence in support of this claim is presented in Sec.~\ref{sec:jrkky}, 
where we evaluate the effective magnetic exchange interaction.

\begin{table*}
\caption{Relevant effective magnetic interaction parameters $J_{ij}/t \ [10^{-4}]$. 
In the upper table the long-range interactions 
$J_1$, $J_2$, $J_3$ denote exchange between edge $C_e^{2}$ atoms 
belonging to neighboring, next-nearest neighboring, and opposite edges 
of the ZGNF, respectively. 
While $J_1$ and $J_3$ always connect $C_e^{2}$ atoms on different sublattices, 
$J_2$ connects $C_e^{2}$ atoms on the same sublattice, 
and assumes different values if the sublattice symmetry is broken. 
In the lower table the NN interactions denote exchange  
between $C_b^{1}$ pairs, $C_e^{1}$ pairs, or $C_b^{4}$-$C_e^{2}$ pairs. 
The latter assumes different values if the sublattice symmetry is broken. 
The above magnetic exchanges are also indicated graphically 
on the ZGNF in Fig.~\ref{fig:jrkky}, for the sake of clarity.}
\label{tab:jrkky_src}
\vspace{0.1cm}
\begin{tabular}{ccccc|cccc}
\hline \hline
\multicolumn{1}{c}{ } & \multicolumn{4}{c|}{$T/t=0.005 \ (T\approx 80$~K)} & \multicolumn{4}{c}{$T/t=0.010 \ (T\approx 160$~K)} \\
\hline \hline
\enskip \enskip \enskip & \enskip \enskip $J_1^{\cal AB}$ \enskip \enskip & \enskip \enskip $J_2^{\cal AA}$ \enskip \enskip & \enskip \enskip $J_2^{\cal BB}$ \enskip \enskip & \enskip \enskip $J_3^{\cal AB}$ \enskip \enskip \enskip & \enskip \enskip \enskip $J_1^{\cal AB}$ \enskip \enskip & \enskip \enskip $J_2^{\cal AA}$ \enskip \enskip & \enskip \enskip $J_2^{\cal BB}$ \enskip \enskip & \enskip \enskip $J_3^{\cal AB}$ \enskip \enskip \\
\hline
$\delta=0$ &  0.38(5) &  0.01(6) & 0.01(6) &  0.04(5) &   0.38(4) & 0.01(6) & 0.01(6) & 0.04(5) \\ 
$\delta=1$ & -0.34(9) &  0.26(5) & 0.17(1) & -1.08(5) & -0.21(9) & 0.14(3) & 0.14(5) & -0.75(7) \\
$\delta=2$ &  0.05(9) &  0.02(4) & 0.02(5) &  0.09(1) & -0.05(0) & 0.00(6) & 0.01(8) & -0.08(6) \\ 
$\delta=3$ &  0.03(9) & -0.01(1) & 0.01(5) & -0.00(6) & -0.01(4) & -0.00(1) & -0.00(1) & -0.00(1) \\ 
\hline 
\\
\hline \hline
\enskip \enskip \enskip & \enskip \enskip $J_{\rm NN}^{\cal AB}$ $C_b^{1}$-$C_b^{1}$ \enskip & \enskip $J_{\rm NN}^{\cal AB}$ $C_e^{1}$-$C_e^{1}$ \enskip & \enskip $J_{\rm NN}^{\cal AB}$ $C_b^{4}$-$C_e^{2}$ \enskip & \enskip $J_{\rm NN}^{\cal BA}$ $C_b^{4}$-$C_e^{2}$ \enskip & \enskip $J_{\rm NN}^{\cal AB}$ $C_b^{1}$-$C_b^{1}$ \enskip & \enskip $J_{\rm NN}^{\cal AB}$ $C_e^{1}$-$C_e^{1}$ \enskip & \enskip $J_{\rm NN}^{\cal AB}$ $C_b^{4}$-$C_e^{2}$ \enskip & \enskip $J_{\rm NN}^{\cal BA}$ $C_b^{4}$-$C_e^{2}$ \enskip \enskip \\
\hline
$\delta=0$ &  1.00(4) &  5.52(4) & 2.70(2) &  2.70(9) & 1.01(3) &  5.47(1) & 2.68(4) &  2.69(3) \\ 
$\delta=1$ &  0.26(2) &  2.05(4) & 0.77(0) &  1.03(1) & 0.27(0) &  2.01(2) & 0.91(6) &  0.91(0) \\
$\delta=2$ &  0.00(3) &  0.02(5) & 0.22(9) & -0.08(1) & 0.02(6) &  0.26(8) & 0.13(7) &  0.07(6) \\ 
$\delta=3$ & -0.00(1) & -0.00(6) & -0.02(0) & 0.04(0) & 0.00(7) &  0.07(2) & 0.02(7) &  0.03(1) \\ 
\hline \hline
\end{tabular}
\end{table*}

\subsection{Effective magnetic interaction}
\label{sec:jrkky}
In the following we analyze the effective magnetic interactions  
generated by the interplay of the local repulsion $U$ and the delocalization 
of electrons and holes in the ZGNF, 
which we argue to be the mechanism behind the stabilization of the FI state. 
Within the local self-energy approximation (as in DMFT), 
an estimate of the effective magnetic exchange interaction parameters $J_{ij}$ 
can be obtained, following Katsnelson and Lichtenstein,\cite{katsnelsonPRB61} as
\begin{equation} \label{eq:jrkky}
 J_{ij} = - \int_{\infty}^{\infty} \! d\omega 
            \Sigma^{s}_{i}(\omega) G^{\uparrow}_{ij}(\omega) 
            \Sigma^{s}_{j}(\omega) G^{\downarrow}_{ji}(\omega) 
            f(\omega), 
\end{equation}
where $f(\omega)={\big(e^{\beta(\omega-\mu)}+1\big)}^{-1}$ 
is the Fermi distribution function at the inverse temperature $\beta=1/T$, 
while $\Sigma^{s}_{i} = ( \Sigma^{\uparrow}_{i} - \Sigma^{\downarrow}_{i} )/2$ 
is the asymmetric spin combination of the local (dynamical) self-energy, 
and $G^{\sigma}_{ij}$ is the real-space non-local Green's function 
connecting sites $i$ and $j$, with spin $\sigma$. 
Diagrammatically, the effective exchange $J_{ij}$ 
can be thought as the frequency convolution 
of the bubble term $\chi^0_{ij}$ of the non-local susceptibility, 
with the non-local Green's function as the fermionic lines of the bubble 
and $\Sigma^{s}_{i}$ playing the role of the local vertex. 
Let us stress that the coupling $J_{ij}$ is zero in the PM state 
(where $\Sigma^{\uparrow}_{i}=\Sigma^{\downarrow}_{i}$) 
and it should not be interpreted as the magnetic coupling 
of an effective spin lattice Hamiltonian (e.g., of the Heisenberg model), 
as it carries a temperature and doping dependence 
through both the Green's function and self-energy. 
Rather, Eq.~(\ref{eq:jrkky}) resembles 
the typical expression\cite{black-schafferPRB81,uchoaPRL106} 
used to evaluate the RKKY exchange coupling between magnetic adatoms:  
$J_{\rm RKKY} \propto J^2 \chi^0_{ij}$, 
where $J$ couples the impurity spin with the spin density on the substrate, 
and $\chi^0_{ij}$ is the static spin susceptibility of the conduction electrons, 
which mediate the effective magnetic interaction. 
In analogy, we have that $\Sigma^{s}_{i}$ is associated with the presence 
of localized magnetic moments and the effective exchange between them 
is mediated by the doped holes delocalized on the lattice. 

\begin{figure}[b]
\includegraphics[width=0.45\textwidth, angle=0]{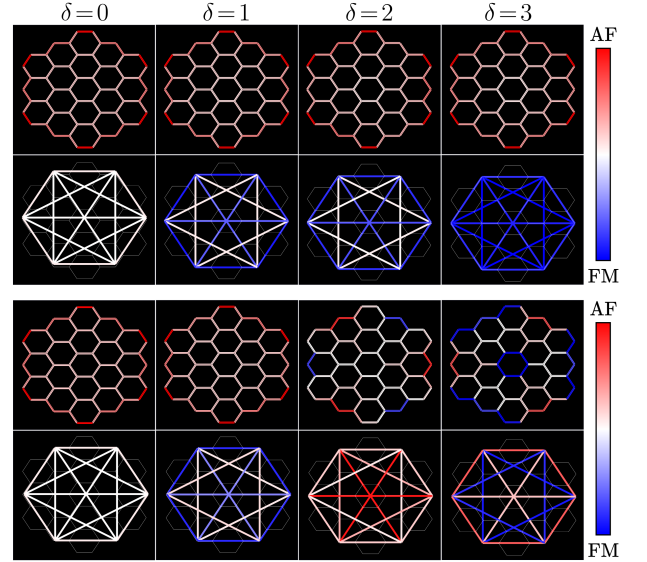}
\caption{(Color online) Effective magnetic exchange $J_{ij}$ 
between all NN pairs and edge $C_e^{2}$ atom pairs for different dopings. 
The color and the intensity of the links denote the nature and the strength 
of the magnetic interaction: from AF (red/light grey) to FM (blue/dark grey). 
The data are normalized to the strongest $J_{ij}$ for a better visibility,  
while the numerical values of the couplings 
are given in Tab.~\ref{tab:jrkky_src} as a reference.
[Upper panels] AF state, at $T/t=0.010$ ($T \approx 80$~K). 
[Lower panels] FI state, at $T/t=0.005$ ($T \approx 160$~K). }
\label{fig:jrkky}
\end{figure}

\begin{figure}[b]
\includegraphics[width=0.47\textwidth, angle=0]{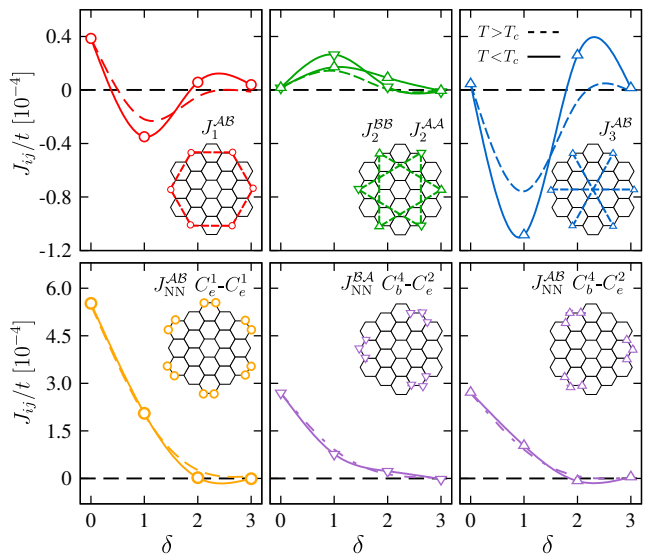}
\caption{(Color online) Effective parameters for 
relevant magnetic interactions $J_{ij}$ as a function of doping $\delta$. 
The dashed and solid lines (guides to the eye) with open symbols 
correspond to $T/t=0.010$ ($T>T_c$) and $T/t=0.005$ ($T<T_c$), respectively. 
[Upper panels] Long-range interactions $J_1$, $J_2$, $J_3$. 
[Lower panels] $J_{NN}$ interactions 
between $C_e^{1}$-$C_e^{1}$, and $C_b^{4}$-$C_e^{2}$ pairs.  
The corresponding values are given in Tab.~\ref{tab:jrkky_src}. } 
\label{fig:jeff_ij_doping}
\end{figure}

The expression for $J_{ij}$ in Eq.~(\ref{eq:jrkky}) yields 
the effective magnetic exchange for any pair $(i, j)$ in the ZGNF. 
However, in order to highlight the microscopic mechanism 
behind the AF-to-FI magnetic transition,  
we focus on the magnetic exchange parameters 
between all NN atom pairs and all edge $C_e^{2}$ atom pairs. 
In Fig.~\ref{fig:jrkky} we show a graphical representation on the ZGNF 
of the values obtained for relevant magnetic exchange parameters $J_{ij}$. 
We show data as a function of doping and for two values of temperature, 
$T/t=0.010$ ($T \approx 160$~K) and $T/t=0.005$ ($T \approx 80$~K), 
which are representative of the high-$T$ (AF) and low-$T$ (FI) 
magnetic states discussed above. 
At half-filling, we find all magnetic interactions $J_{ij}>0$ (i.e., AF in nature). 
The short-range, and in particular the NN interactions $J_{NN}$ 
are stronger at the edges with respect to the bulk. 
The values of $J_{ij}$ are rapidly suppressed with distance $|i-j|$, 
and in particular long-range interactions between different edges 
are negligibly weak with respect to $J_{NN}$ within a given edge. 
The magnetic properties at half-filling are weakly dependent 
on $T$ (below $T_N$) and this is reflected also in the magnetic couplings. 
At finite doping and above $T_c$ we find $J_{NN}>0$ 
and quantitatively similar to the ones at $\delta=0$ for all values of doping. 
However, the presence of delocalized charge carriers 
mediate sizable long-range magnetic interactions. 
Some of the long-range interactions connecting edge $C_e^{2}$ atoms, 
indicated as $J_1$, $J_2$, and $J_3$ in Tab.~\ref{tab:jrkky_src}, 
are found to be negative (i.e., FM in nature) at finite doping. 
Eventually, the presence of $J_{ij}<0$ drives 
the onset of the FI state as $T \rightarrow T_c(\delta)$. 
Interestingly, below $T_c(\delta)$ 
the change in the magnetic structure at $\delta \neq 0$ 
is reflected also in a change of the effective exchange interactions. 
The results are clearer for $\delta=2$ and $\delta=3$, 
where an exact correspondence can be found between 
the $J_{NN}$ shown in Fig.~\ref{fig:jrkky} 
and the relative orientation of the corresponding pair 
of magnetic moments shown, e.g., in the insets 
of Figs.~\ref{fig:awPM2AFn0963Tf} and \ref{fig:awPM2AFn0945Tf}, respectively. 
The behavior of long-range interactions is less obvious 
and better illustrated in Fig.~\ref{fig:jeff_ij_doping}, 
in which we compare the doping dependence 
of both representative $J_{NN}$ and the $J_{1-3}$ 
magnetic exchange interactions above and below $T_c$. 
In general $J_{NN}$ become weaker upon doping, 
with some of them (in particular at the ZZ edges) 
becoming FM at $\delta=2$ and $\delta=3$ below $T_c$. 
Instead, long-range interactions 
are enhanced at $\delta=1$ with respect to the half-filling case, 
but are suppressed upon further increasing the doping. 
The interaction $J_1$ and $J_3$, that connect edge $C_e^{2}$ atoms 
of different sublattices, are usually larger than $J_2$  
and display an oscillatory behavior, changing sign as a function of doping. 
In particular, the FM nature of $J_1<0$ and $J_3<0$ above $T_c$ 
reveals the tendency of magnetic moments at the ZZ edges to align FM, 
and can be interpreted as the microscopic mechanism 
driving the system a cross the AF-to-FI transition.  
Below $T_c$, we find $J_{1-3}>0$ for $\delta=2$ and $\delta=3$ 
and we interpret it as a signature of the competition between the AF and the FI states. 
Evidence for the coexistence and the cooperation of AF and FM correlations 
in determining the magnetic state of doped triangular and linear chain ZGNFs 
was already discussed by Chacko {\it et al.}\cite{chackoPRB90} 
within exact diagonalization calculations. 
This hints at the generality of the above scenario 
in graphene nanostructures, which not limited to a particular shape, 
but seems to be is a general feature related to the presence of ZZ edges. 
Note, however, that the values of the FM $J_{ij}$ couplings 
extracted in the calculations are relatively weak 
compared to the AF $J_{NN}$ and the temperature scale $T_c$ 
at which the FI magnetic order sets in. 
This suggests that also the geometry of the ZGNF 
plays an important role, assisting the exchange couplings 
in the formation of the FI state.

\section{Summary and Outlook} 
\label{sec:summary}
In this work we investigated the interplay 
between charge and spin degrees of freedom in the magnetic properties 
of a doped ZGNF within the framework of DMFT. 
At half-filling we analyze the onset of magnetism 
as a function of the local interaction $U$. 
We identify a dichotomy between bulk and edge C atoms, 
which persist from weak- to strong-coupling. 
Above a threshold value of $U$, the ground state of the ZGNF 
is in a fully-compensated AF state. 
The analysis of the energy balance within DMFT 
suggests that, for realistic values of the interaction, 
at half-filling the AF state is stabilized by a weak-coupling mechanism. 
The results obtained are in qualitative agreement 
with static-MFT and DFT calculations, 
but show that quantum fluctuations suppress AF 
with respect to mean-field approximations. 

Upon introducing charge carriers we observe the melting of the AF state. 
Below a doping-dependent ordering temperature $T_c(\delta)$ 
it is possible to stabilize a short-range FI order, 
in which the magnetic moments at the ZZ edges are aligned FM. 
In the FI state the ZGNF displays a net ferromagnetic moment 
which coexists with a finite staggered magnetization.
We interpret the change in the magnetic configuration 
in terms of an effective magnetic exchange between the ordered spins, 
mediated by the charge carriers localized in the proximity of the edges. 

The possibility of driving FM correlations upon doping 
was already discussed in the framework of DFT\cite{kabirPRB90} 
and exact diagonalization.\cite{chackoPRB90} 
The overall agreement with these studies indicates 
that a reasonable description of the magnetic phases can already be obtained 
relying on a mean-field description of long-range correlations. 
In this framework, DMFT has allowed us to accurately capture the interplay 
between the incoherent excitations that form the fluctuating local moment 
and the coherent low-energy excitations 
that screen this local moment on longer time scales 
and mediate the magnetic exchange which stabilize the ordered state. 
Moreover, the possibility of describing both the temperature and doping dependence 
of the effective exchange couplings sheds some light 
on the onset of the FI short-range order 
and the strong competition between AF- and FM-correlations in ZGNFs. 

Evidently, any change in the magnetism and in the low-energy spectral properties 
will have important consequences on the transport through ZGNFs. 
Hence, the above analysis indicate the electrostatic control 
of the magnetization of doped ZGNF as a promising route 
towards the future conception and realization of carbon based spintronic devices.

\begin{acknowledgments}
We acknowledge valuable discussions 
with R.~Drost, D.~Prezzi, and Z.~Zhong. 
We are grateful to G.~Sangiovanni for helpful discussions 
in several stages of this work, and to M.~Pickem who provided 
a benchmark for the static-MFT results. 
We acknowledge financial support from the Austrian Science Fund (FWF) 
through the Erwin~Schr\"{o}dinger fellowship J3890-N36 (AV), 
through project I-610-N16 (AT), and the SFB ViCoM (AT, AV, KH), 
as well as the European Research Council under the European Union's 
Seventh Framework Program FP7/ERC through grant agreement 
n.~306447 (KH, AV), n.~240524 (MC, AA, AV), and n.~280555 (AA). 
\end{acknowledgments}

\appendix

\section{Details of the numerical simulations}
In the following we discuss the technical details 
of the magnetic real-space DMFT calculations 
and the annealing procedure used to obtain the temperature evolution 
of the magnetic properties of the ZGNF. 

The auxiliary AIMs of the real-space DMFT algorithm 
are solved with a L\'{a}nczos exact diagonalization 
impurity solver\cite{caffarelPRL72,caponePRB76} 
which is able to accurately describe the physics both at $T=0$ and at finite $T$. 
We employ a typical discretization of the Hilbert space of $n_s = 1 + n_b = 9$ sites, 
with $n_b$ being the number of bath sites connected to the impurity. 
In specific cases we also performed calculations up to $n_s=12$ sites, 
finding no qualitative difference in the physical observables. 
The reliability of the finite temperature results 
obtained with L\'{a}nczos exact diagonalization impurity solver 
was tested against continuous-time Quantum Monte Carlo impurity solver 
implemented in the w2dynamics package,\cite{parraghPRB86} 
showing quantitative agreement in the physical observables.

In order to get a magnetic solution within real-space DMFT, 
we lift the local $SU(2)$ spin rotational symmetry of each auxiliary AIM. 
The symmetry is manually broken at the beginning of the self-consistency cycle  
by applying a symmetry-breaking filed $\eta_{i\sigma}$ 
to the spin-dependent DMFT bath ${\cal G}_{0i\sigma}(\omega)$. 
In the case of a AF state, symmetry-breaking field takes the form 
\begin{equation} \label{eq:sbf}
 \eta_{i\sigma} = \begin{cases} \eta (\delta_{\sigma\uparrow}-\delta_{\sigma\downarrow}),
                                & \mbox{if } i \in {\cal A} \\
                                \eta (\delta_{\sigma\downarrow}-\delta_{\sigma\uparrow}), 
                                & \mbox{if } i \in {\cal B}
                  \end{cases}
\end{equation}
where we set the parameter $\eta = 0.05t > 0$. 
The field in Eq.~(\ref{eq:sbf}) corresponds to a staggered 
perturbation with the same symmetry of the fully-compensated AF state. 

At half-filling the system is unstable toward AF 
and the convergence of the DMFT self-consistency is smooth down to $T=0$. 
At finite doping $\delta \neq 0$, 
besides the solution of the inhomogeneous real-space DMFT equations, 
one also needs to determine the chemical potential $\mu(n)$ 
corresponding to electron concentration 
$n = \sum_{i\sigma} \langle n_{i\sigma} \rangle$. 
The search for $\mu(n)$ involves a complex root-finding 
within the self-consistent procedure, 
which makes the convergence of the DMFT self-consistency 
numerically unstable for arbitrary values of $T$ and $\delta$. 
The difficulty of the root-finding is also enhanced 
due to the discreetness of the energy spectrum for a nanoscopic system. 
However, motivated by physical observations, 
it is possible to obtain a reliable self-consistent solution 
of the DMFT equations in a wide range of $T$ and $\delta$ 
by following an annealing procedure. 
Indeed, one can notice that at high-$T$, 
AF short-range magnetic correlations are dominant, 
as evident by the values of the effective magnetic exchange $J_{ij}$ 
shown in Tab.~\ref{tab:jrkky_src}.  
The AF correlations stabilize a fully-compensated AF state, 
also at $\delta \neq 0$. 
A staggered spatial order of the magnetic moments 
can be easily obtained at high-$T$ 
with the natural choice in Eq.~(\ref{eq:sbf}) for $\eta_{i\sigma}$, 
mainly for two reasons: 
(i) the magnetic ground state displays a spatial distribution 
of the magnetic moments that closely resembles the initial state 
given by the symmetry-breaking field $\eta_{i\sigma}$; 
(ii) the temperature broadening soothes the complexity of the root-finding 
in the case of a discrete energy spectrum. 
Once a high-$T$ calculation is converged, 
the chemical potential $\mu$, 
the spin-dependent Weiss fields ${\cal G}_{0i\sigma}(\omega)$, 
and the list of the L\'{a}nczos states for each inequivalent atom 
are used as an input for the calculation at lower $T$. 
As the input Weiss field for the next calculation 
is already symmetry-broken, the annealing procedure in continued \emph{without} 
imposing a symmetry-breaking field $\eta \neq 0$ with any specific spatial structure. 
This way, we observe a \emph{spontaneous} transition 
toward the FI state in the low-$T$ regime.  
In analogy, a reverse annealing procedure was followed 
starting from a converged low-$T$ calculation. 
This allowed to reveal the hysteretic behavior 
of the magnetic moments at $\delta \neq 0$.


\begin{thebibliography}{99}
\bibitem{note:doping} With a particle-hole symmetric Hamitonian 
the physics does not change upon electron doping. 
However, this is no longer true if structural relaxations for the non-passivated ZGNF 
are considered.
\end{thebibliography}
\end{document}